\documentclass[preprint,10pt]{aastex}

\newcommand{\sfrunit}{\msol $\mbox{yr}^{-1}$}

\newcommand{\FBDB}{{\tt Run~S1}}      
\newcommand{\FBDT}{{\tt Run~S2}}      
\newcommand{\FBDG}{{\tt Run~SG1}}       
\newcommand{\FBDA}{{\tt Run~SF1}}      
\newcommand{\FBDR}{{\tt Run~SF2}}      
\newcommand{\EAAJ}{{\tt Run~Ha}}      
\newcommand{\EACB}{{\tt Run~Hb}}      
\newcommand{\sFBDB}{{\tt S1}}
\newcommand{\sFBDT}{{\tt S2}}
\newcommand{\sFBDG}{{\tt SG1}}
\newcommand{\sFBDA}{{\tt SF1}}
\newcommand{\sFBDR}{{\tt SF2}}
\newcommand{\sEAAJ}{{\tt Ha}}
\newcommand{\sEACB}{{\tt Hb}}

\newcommand {\has} {H$\alpha$}
\newcommand{\BB}{\textit{\texttt{B}\texttt{\has}\texttt{BAR}\,\,}}
\newcommand{\FM}{\texttt{\textsc{FaNTOmM}}}
\newcommand {\hI} {\ion{H}{1}\,\,}

\newcommand {\ha} {H$\alpha$\,\,}
\newcommand {\kms} {\,km\,s$^{-1}$\,}
\newcommand {\M} {\mbox{${\cal M}$}}

\newcommand {\msol} {\M$_\odot$\,}

\newcommand{\Min}{${}^{\prime}$}
\newcommand{\Sec}{${}^{\prime\prime}$}
\newcommand{\Deg}{${}^{\circ}$}

\newcommand{\Arc}{''}

\slugcomment{To be published to {\it Astrophysical Journal}}
\shorttitle{Pattern speeds determination in M100}
\shortauthors{Hernandez et al.}

\begin{document}

\title{On the relevance of the Tremaine-Weinberg method applied to H$\alpha$ velocity field.Pattern speeds determination in M100 (NGC 4321).}

\author{Olivier Hernandez}
\affil{D\'epartement de physique and Observatoire du mont
M\'egantic, Universit\'e de Montr\'eal, C.P. 6128, Succ. centre
ville,Montr\'eal, Qu\'ebec, Canada. H3C 3J7}
\email{olivier@astro.umontreal.ca}
\author{Herv\'e Wozniak}
\affil{Centre de Recherche Astronomique de Lyon, 9 avenue Charles Andr\'e,
F-69561 Lyon, France}
\email{herve.wozniak@obs.univ-lyon1.fr}
\author{Claude Carignan}
\affil{D\'epartement de physique and Observatoire du mont
M\'egantic, Universit\'e de Montr\'eal, C.P. 6128, Succ. centre
ville, Montr\'eal, Qu\'ebec, Canada. H3C 3J7}
\email{carignan@astro.umontreal.ca}
\author{Philippe Amram}
\affil{Observatoire Astrophysique Marseille Provence, Laboratoire
d'Astrophysique de Marseille,2 Place Le Verrier, F--13248  Marseille Cedex 04, France}
\email{philippe.amram@oamp.fr}
\author{Laurent Chemin}
\affil{D\'epartement de physique and Observatoire du mont
M\'egantic, Universit\'e de Montr\'eal, C.P. 6128, Succ. centre
ville,Montr\'eal, Qu\'ebec, Canada. H3C 3J7}
\email{chemin@astro.umontreal.ca}
\and
\author{Olivier Daigle}
\affil{D\'epartement de physique and Observatoire du mont
M\'egantic, Universit\'e de Montr\'eal, C.P. 6128, Succ. centre
ville,Montr\'eal, Qu\'ebec, Canada. H3C 3J7}
\email{odaigle@astro.umontreal.ca}

\clearpage

\begin{abstract}
The relevance of the Tremaine-Weinberg (TW) method is tested to
measure the bar, spiral and inner structure pattern speeds using a
gaseous velocity field. The TW method is applied to various
simulated barred galaxies in order to demonstrate its validity in
seven different configurations, including star formation or/and
dark matter halo. The reliability of the different physical
processes involved and of the various observational parameters are
also tested. The simulations show that the TW method could be
applied to the gaseous velocity fields to get a good estimate of
the bar pattern speed, under the condition that regions of shocks
are avoided and measurements are confined to regions where the
gaseous bar is well formed. We successfully apply the TW method to
the \ha velocity field of the Virgo Cluster Galaxy M100 (NGC 4321)
and derive pattern speeds of $55\pm5$ \kms kpc$^{-1}$ for the nuclear
structure, $30\pm2$ \kms kpc$^{-1}$ for the bar and $20\pm1$ \kms kpc$^{-1}$ for the spiral pattern, in full agreement with published determinations
using the same method or alternative ones.
\end{abstract}


\section{Introduction}

The presence of a barred structure appears to be a common
attribute of disk galaxies. Roughly 30\% of spiral galaxies are
strongly barred in the optical (de Vaucouleurs 1963) while another
25\% are weakly barred.
Evidences that bars in spirals are more obvious in the
near-infrared (NIR) than in the visible go back to Hackwell \&
Schweizer (1983).  More recent surveys in the NIR have shown that
up to 75\% of high surface brightness galaxies may have a more or
less strong bar (e.g. Knapen, Shlosman \& Peletier 2000;
Eskridge et al. 2000).
Strong bars are nearly twice as prevalent in the NIR
than in the optical.
The high frequency of occurence of bars means that they are
long-lived attributes of disk systems.
While bars should be destroyed rapidly, it is tought
that continuous accretion produces multi-phase bars (Bournaud \& Combes 2002) .

The fact that bars may contain a large mass fraction of the disk
suggests that they are a fundamental component of the mass
distribution in spiral galaxies. Since their kinematics is
different from the one of the more or less axisymmetric disk, it
is important to model them properly to derive, as accurately as
possible, the overall mass distribution. This is especially
important since bars are in the inner parts of disk systems which
is the region where the free parameters of the mass models are
best constrained (Blais-Ouellette et al. 1999, 2004; Blais-Ouellette,
Amram \& Carignan 2001). The
parameters of mass models are not constrained by the flat part but
by the rising part of rotation curves (RC).

The determination of the bar pattern speed ($\Omega_p$) in spiral
galaxies is one of the most important kinematical parameters since
it drives a large part of their evolution (e.g. Block et al. 2004). It is essential for understanding the so-called dark matter
problem and, more generally, the structure of spiral galaxy halos.

A determination of the bar pattern speed and other
non-axisymmetric and asymmetric structures is important for a
number of reasons:
\begin{itemize}
    \item Orbital calculations show that a bar can only be built
self-consistently if it lies entirely within its corotation
radius; the point at which the bar pattern rotates at the
same speed as a star on a circular orbit at that radius
 (Contopoulos 1980). Further studies using hydrodynamical
simulations have shown that the corotation radius has to lie in
the interval of 1.2$\pm$0.2 length of the semi-major bar axis
(Athanassoula 1992a).
    \item Bars may initiate spiral density waves (e.g. Toomre 1969)
and stellar rings (e.g. Buta 1986). A clear determination of the
corotation may give insight on the nature of density waves and on
the energy and angular momentum exchanges between the bar and the
disk (Sellwood 1985; Masset $\&$ Tagger 1997).
    \item Stellar bars may provide a means to transport gas towards
the nuclear regions of galaxies by inducing gravitational torques
(Block et al. 2002) and fuel nuclear starbursts or active galactic
nuclei (e.g. Kormendy 1982).  The transfer of material (inflow or
outflow) may be linked to the pattern speed (Athanassoula 1992b).
    \item Bars may drive the secular evolution of bulges,
by triggering nuclear starbursts and by kinematic heating of the
inner disk (e.g. Combes et al. 1990).  The bar pattern speed
depends critically on the relative bulge mass and the disk
scale-length  (e.g. Combes \& Elmegreen 1993).
    \item Bars' pattern speeds may help to discriminate (Noguchi 2004)
between spontaneous bars (large $\Omega_p$) and tidal bars (small
$\Omega_p$).
    \item Bars may introduce errors in determining the mass distribution of
spiral galaxies from their rotation curves. Indeed, non-circular
motions are usually azimuthally averaged and may
have for consequence to shallow the shape of the inner regions of
the RCs (e.g. Swaters et al. 2003). These non-circular
motions must be modeled properly and taken out from the observed kinematics
in order to retrieve the true circular kinematics which trace
the mass distribution.
    \item Strongly barred galaxies may have maximal (or nearly maximal)
disks.  The absence of slow bars may require maximum disks, which
transfer very little angular momentum to the low density halos
(Debattista \& Sellwood 1998).  Alternatively, the strongest bars
and higher pattern speeds may be found in sub-maximum disks where
the amount of angular momentum exchanged by resonant particles
between the disk and the halo has been the largest (Athanassoula 2003). Bars are often associated with rings (e.g., Buta 1995) and
may require a maximal disk for stability (e.g. Quillen \& Frogel
1997).

\end{itemize}

The TW method is described in Sect. 2, tests on numerical
simulations are presented in Sect. 3, the case of M100 is examined
in Sect. 4 and the discussion and conclusions are given in Sect. 5 and 6.

\section{Description of the Method}

Bar pattern speeds of spirals are determined by identifying
theoretically predicted resonances (Lindblad resonances,
corotation) with periodic motions of the stars and gas (extracted
from the RC). The determination of bar pattern speeds relies on
observational methods or, alternatively, on matching numerical
models (N-body + SPH codes, see next section) to the observed
velocity fields. A number of reviews describing these methods in
detail can be found (e.g. Teuben 2002).

Two methods, somewhat indirect, are based on the identification of
morphological features associated with resonance radii (measuring
the inner resonance 4:1, Elmegreen, Elmegreen \& Montenegro 1992;
or alternatively, measuring the sign inversion of the radial
streaming motions across corotation, Canzian 1993). Theoretical
evidences argue that the corotation radius lies just beyond the
end of the bar (e.g. pioneer work of Contopoulos 1980).  Based on
this result, many bar pattern speeds have been estimated (see a
compilation in Elmegreen et al. 1996).

A more direct model independent method, not relying on any
particular theory of density waves, is due to Tremaine \& Weinberg
(1984, hereafter TW). The pattern speed is determined from two
observationally accessible quantities: the luminosity-weighted
mean velocities and the luminosity-weighted mean densities
throughout the disk of the galaxy.  In the plane of the sky, X is
the coordinate along the major axis of the galaxy, Y along the
minor axis and $V_{LOS}$(X,Y) the line of sight velocity. $<X_Y>$
is the luminosity-weighted mean X-position integrated along a
strip parallel to the X-axis at Y coordinate, $<V_{LOS,Y}>$ is the
luminosity-weighted mean velocity along the X-axis at Y coordinate.
Thus, $$
\Omega_p \sin i =\frac{ \int_{-\infty}^{\infty}\Sigma(x,y,t)V_y(x,y,t)\,dx}{ \int_{-\infty}^{\infty}\Sigma(x,y,t)x(y,t)\,dx} = \frac{<V_{LOS,Y}>}{<X_Y>},
$$
where $\Omega_p$ is the pattern speed and $\Sigma$ is the surface density of the
component.

The underlying assumption is therefore that the density is proportional to intensity. Therefore, a corollary statement is that disks must be close to maximum in the inner parts for the method
to work. The TW method also requires a tracer population that
satisfies the continuity equation (no significant creation or
destruction of matter over an orbit). It may be the case for the
stellar population, even if, in reality, the continuity equation
is never strictly satisfied because of the continuous star
formation. Nevertheless, old stars may survive many passage
through the pattern.  As long as the star formation efficiency is
low, conversion of gas into stars (and vice versa through winds and supernovae) can be ignored

For these reasons, the TW method has preferentially been applied
to early-type bars using starlight and absorption-line kinematics
from long-slit spectra: NGC936 (Kent 1987; Kent \& Glaudell 1989;
Merrifield \& Kuijken, 1995); NGC4596 (Gerssen, Kuijken \&
Merrifield 1999); NGC1023 (Debattista, Corsini \& Aguerri 2002); NGC7079
(Debattista \& Williams 2001); ESO 139-G009, IC874, NGC1308,
NGC1440 and NGC3412 (Aguerri, Debattista \& Corsini 2003); NGC271,
NGC1358, ESO 281-31, NGC3992 (Gerssen, Kuiken \& Merrifield 2003).

Application of the TW method to gaseous phases is complicated
especially since the method assumes that the disk component obeys the
continuity equation  and that the relation between the emission
and its surface density is linear (or can be modeled). Atomic
hydrogen and molecular gas will not obey the continuity equation,
because of the conversion of gas between the three phases (molecular,
atomic and ionized) and
because the star formation timescales are shorter
than the orbital timescales.
However, Zimmer, Rand, \& McGraw (2004) and Rand
\& Wallin (2004) have argued that, in galaxies in which either the
\hI or H$_2$ everywhere dominates the large
majority of the ISM, the conversion processes can be neglected
and the TW method applied.

Pattern speeds using the TW method were recently derived using \hI
for M 81 by Westpfahl (1998) and for NGC 2915 by Bureau et al
(1999). Using CO observations, pattern speeds were also determined
in galaxies with molecule-dominated ISMs for M51, M83, and N6946
by Zimmer, Rand, \& McGraw (2004) and for NGC1068, NGC3627, NGC4321
(M100), NGC4414, NGC4736 and NGC4826 (from the BIMA SONG survey) by Rand
\& Wallin (2004).

Gas in its ionized phase will never
dominate the ISM; the equation of continuity for the ionized gas
will never be satisfied on an orbital period and the \ha
luminosity is not supposed to trace the mass.
HII regions are bright ionized regions
surrounding massive and hot newborn O and B stars. The OB stars
have main-sequence lifetime of only a few $10^6$ years,  HII
regions are furthermore short-lived gas clouds (which only exist
during the lifetime of their ionizing OB stars) embedded in a
star-forming region in a molecular cloud.  A given HII region,
associated to its parent stars, cannot survive to a typical
galactic rotation of 10$^{8}$ years. Nevertheless, HII regions and
OB stars have not enough time to wander far from their parent
molecular cloud and are, if we neglect the expansion processes, a
tracer of the molecular gas density.
Moreover, $<V_{LOS,Y}>$ determined from \has, is an as good
tracer as any of the galactic potential well (as it is the case when
deriving a RC).

It is true to say that the application
of the TW method to \ha may apparently violate several of the TW
conditions. However, we suppose that, to first order, the continuity
equation is satisfied, but only for very short times ($\ll$~orbital
period), we neglect the internal kinematics of HII regions and we
suppose that the luminosity-weighted mean X-position integrated
along a strip parallel to the X-axis, $<X_Y>$, is a valuable indicator
of the mean mass distribution.

Preliminary results on the determination of bar pattern speeds
using HII regions were already presented for NGC2903, NGC3359, NGC4321 (M100),
NGC5194 and NGC6946 (Hernandez et al. 2004).

\section{Tests on Numerical Simulations}

The TW method will be applied to simulated barred galaxies in order to
demonstrate its validity and to understand its sensitivity to
various observational parameters.

\subsection{Description of the simulations}   \label{sec:Simulations}
\subsubsection{The codes}
This section will describe the techniques used to perform
self-consistent simulations including stars, gas and star formation.
In order to check for model and/or code dependent results, 
two significantly different numerical schemes 
were applied for the computation of the forces.

A particle--mesh N-body code was used which includes stars, gas and
recipes to simulate star formation (SF). The broad outlines of the code are
the following: the gravitational forces are computed with a
particle--mesh method using a 3D polar grid with $(N_R, N_\phi,
N_Z)=(31,32,64)$ active cells (Pfenniger \& Friedly 1993), the hydrodynamical equations are solved using the \textsc{SPH} technique (Benz 1990) and the star formation process is based on Toomre's criterion for the radial instability of gaseous discs (see Friedli \& Benz 1995, Michel-Dansac \& Wozniak 2004 for more details). When star formation is active, the radiative cooling of the gas has been
computed assuming solar metallicity.

When an extended live dark matter component was added,
our computations were performed with \texttt{GADGET},  a tree-based
N-body$+$\textsc{SPH} code developed by Springel, Yoshida \& White (2001). An adaptative time step was used,
based on the dynamical time and limited by the Courant condition (Springel, Yoshida \& White 2001).

\subsubsection{Initial conditions}

For all the simulations, an initial stellar population is setup to
reproduce a disc galaxy with an already formed bulge. These particles
form the `old population' as opposed to particles created during the
evolution (`new population') for simulations with star formation
switched on.

For all runs, the initial positions and velocities of the stellar
particles are drawn from a superposition of two axisymmetric
Miyamoto-Nagai discs (Miyamoto \& Nagai 1975) of mass respectively $10^{10}$ and
$10^{11}$\msol, of scale lengths respectively $1$ and
$3.5$~kpc and common scale height of $0.5$~kpc. Velocity
dispersion are computed solving numerically the Jeans equations.
Runs numbered ``2'' have 5 times more particles than those
numbered ``1'' (cf. Table \ref{tabhw1})
For runs with a dissipative component (\sFBDG\, \sFBDA, \sFBDR,
\sEAAJ\ and \sEACB), the gas is represented by 50\,000 particles for a
total mass of $10^{10}$\msol distributed in a disc of scale
length $3$~kpc. For \FBDG\ the gas is kept isothermal while, when the
star formation is switched on, a cooling function is assumed for a
solar metallicity and the energy conservation equation is solved at
each step (Michel-Dansac \& Wozniak 2004).

For the sake of comparison, we have also computed pure N-body models
(\FBDB, \sFBDT), i.e. without gas and star formation. The total mass
is the same mass than the total initial mass of the visible components of
runs \sFBDA, \sFBDR, \sEAAJ\ and \sEACB. Of course, because of star
formation, the stellar mass increases for runs \sFBDA\ and \sFBDR\
while it remains constant for all other runs. These runs tell us what
should be the evolution of the stellar mass and kinematical properties
in the absence of any dissipative component.

Apart from the number of stellar particles (1.1~$\times 10^6$ particles),
the setup of the stellar and gas distributions of runs \sEAAJ\ and
\sEACB\ are  similar to the other runs. \EAAJ\ and \sEACB\ include an
additional dark halo made of 2.2~$\times 10^6$ live particles distributed
in a Plummer sphere of scalelength $50$~kpc and of mass respectively
2.42 and 6.46~$\times 10^{11}$\msol. As our intention is to check the
robustness of our results against the presence of the massive dark
halo, a Plummer sphere is a simple but sufficient description of
the halo. The setup of velocities and velocity dispersions is made
consistent with the dark matter distribution.

%
%
\begin{table*}
\caption[Run parameters]{Run parameters. The masses unit is $10^{11}$\msol. The number of
particles in $10^5$.}
\begin{center}
\begin{tabular}{llllllllll}
\hline \hline 
Model      & \sFBDB & \sFBDG & \sFBDA & \sFBDT & \sFBDR & \sEAAJ & \sEACB \\
\hline
M$_{\mathrm{stars}}$& 1.21   & 1.1    & 1.1    & 1.21   & 1.1    & 1.1    & 1.1    \\
N$_{\mathrm{stars}}$& 5.5    & 5      & 5      & 25.5   & 25     & 11     & 11     \\
M$_{\mathrm{gas}}$  &        & 0.11   & 0.11   &        & 0.11   & 0.11   & 0.11   \\
N$_{\mathrm{gas}}$  &        & 0.5    & 0.5    &        & 0.5    & 0.5    & 0.5    \\
M$_{\mathrm{halo}}$ &        &        &        &        &        & 2.42   & 6.46   \\
N$_{\mathrm{halo}}$ &        &        &        &        &        & 22     & 22     \\
M$_{\mathrm{tot}}$  & 1.21   & 1.21   & 1.21   & 1.21   & 1.21   & 3.63   & 7.67   \\
SF         &        & off    & on     &        & on     & off    & off    \\
\hline
\end{tabular}
\end{center}
\label{tabhw1}
\end{table*}

\subsubsection{Typical evolution of the models}

For all runs without star formation, the initial disc quickly develops
a typical strong bar and a spiral structure both in the stellar and
the gaseous components. The gravity torques due to the bar and the spiral
structure drive the gas inwards and the angular momentum outwards.
The mass distribution is reorganized even for the old stellar
population; this gas inflow occurs on a rather short timescale. The
bar can be unambiguously determined (in size and position-angle) after
300~Myr.
%
%
\begin{figure}
\begin{center}
\plotone{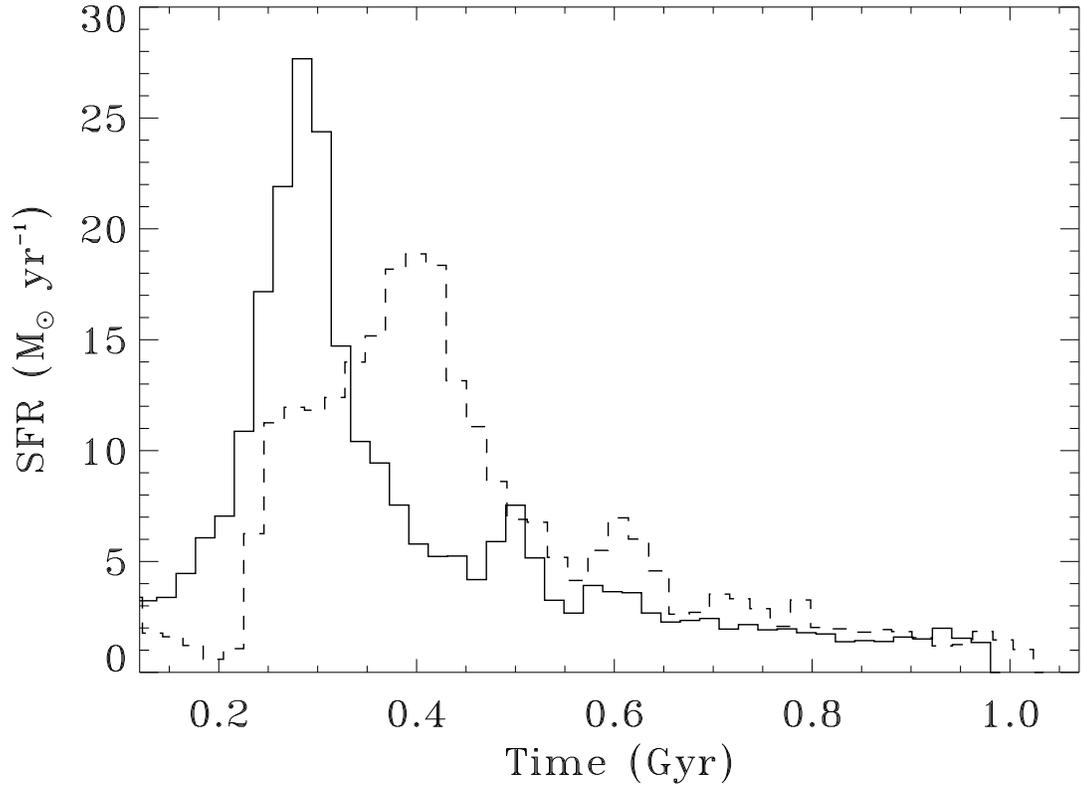}
\end{center}
\caption[Global star formation rate]{Global star formation rate (\sfrunit) versus the elapsed time
from the beginning of \FBDA\ (full line) and \sFBDR\ (dotted
line). Star formation is inhibited during the first 50~Myr for \FBDA\
and 100~Myr for \sFBDR}
\label{fig:SFR}
\end{figure}
For runs with star formation, the global Star Formation Rate (SFR) is displayed in
Figure~\ref{fig:SFR}. For \FBDA\ three SFR maxima occur around 0.3, 0.5
and 0.6~Gyr. For \FBDR, star formation is inhibited during the first
0.1~Gyr in order to delay the SFR peak. Thus two main SFR peaks occur
at $t\sim 0.4$ and $\sim 0.6$~Gyr. The SFR reaches 19--28~\sfrunit.

For $t \la 300$~Myr, the bar and a bi-symmetric spiral structure
spontaneously form as for the runs without star formation. The
gravitational torques on the gas by the bar and the spiral
arms create several regions of very high gas density in which star
formation is ignited. As shown in previous works, during the first
Myrs the SF is not homogeneously distributed over the whole disc but is
mainly concentrated along the bar major axis and along the spiral
arms. After $\sim 1$~Gyr a nuclear gas disk is formed from the
accumulation of gas in the centre and new stars are actively formed
only in this region. The secondary SFR peaks are the result of the gas
inflow towards the central regions of the disc. Indeed, such inflow is
not stationary, but rather proceeds by burst.

Runs \sEAAJ\ and \sEACB\ are stabilized by a live massive dark
matter halo. The formation of the bar is slightly delayed by
0.1~Gyr for \EAAJ ~but after 1~Gyr this run looks like \FBDG. For
\EACB\, the bar appears after $\approx$ 1.5~Gyr. Both runs develop
a bar and a spiral structure. \EACB\ develops an additional inner
ring inside the corotation, which is typical of the interaction of
disc particles with the halo ones (cf. Athanassoula \& Misiriotis
2002).

\subsection{Application of the TW method}

The TW method was applied in an automated way since we are dealing
with numerical simulations. Particle positions and velocities have
been projected to an inclination of 45~$^\circ$. Particles have also
been rotated so that the projected position-angle of the bar with
respect to the line of nodes ($PA_{\mathrm bar}$) lies between 30 and
50~$^\circ$. These are optimal values for a best estimate of
$\Omega_p$ (Merrifield \& Kuijken 1995). For each component (stellar or gas
particles), two (X,Y) frames are computed, one for the mass integrated
along the LOS, another for $V_{LOS}$. The field-of-view is limited to
the projected bar length (between 2 and 5~kpc). The spatial resolution
is 200~pc. The process is repeated for each available snapshots from
$t=0$ to $t=1$~Gyr (2~Gyr in the case of \EACB).

The mass-weighted line-of-sight velocity $<V_{LOS,Y}>$ and the
mass-weighted position coordinate, $<X_Y>$ are computed for each
values of $Y$.  Each $Y$ position thus acts as a long slit. A linear fit
between $<V_{LOS,Y}>$ and $<X_Y>$ is made using a robust least square
algorithm. Since, we exactly know the line-of-nodes position-angle and
the inclination angle, the slope determination of the $<V_{LOS,Y}>$
versus $<X_Y>$ relationship is the only source of errors on
$\Omega_p$. Finally we obtain two estimates of $\Omega_p$, one using
the stellar component $\Omega_p^s$, another with the gas component
$\Omega_p^g$.

\begin{figure}
\begin{center}
\plotone{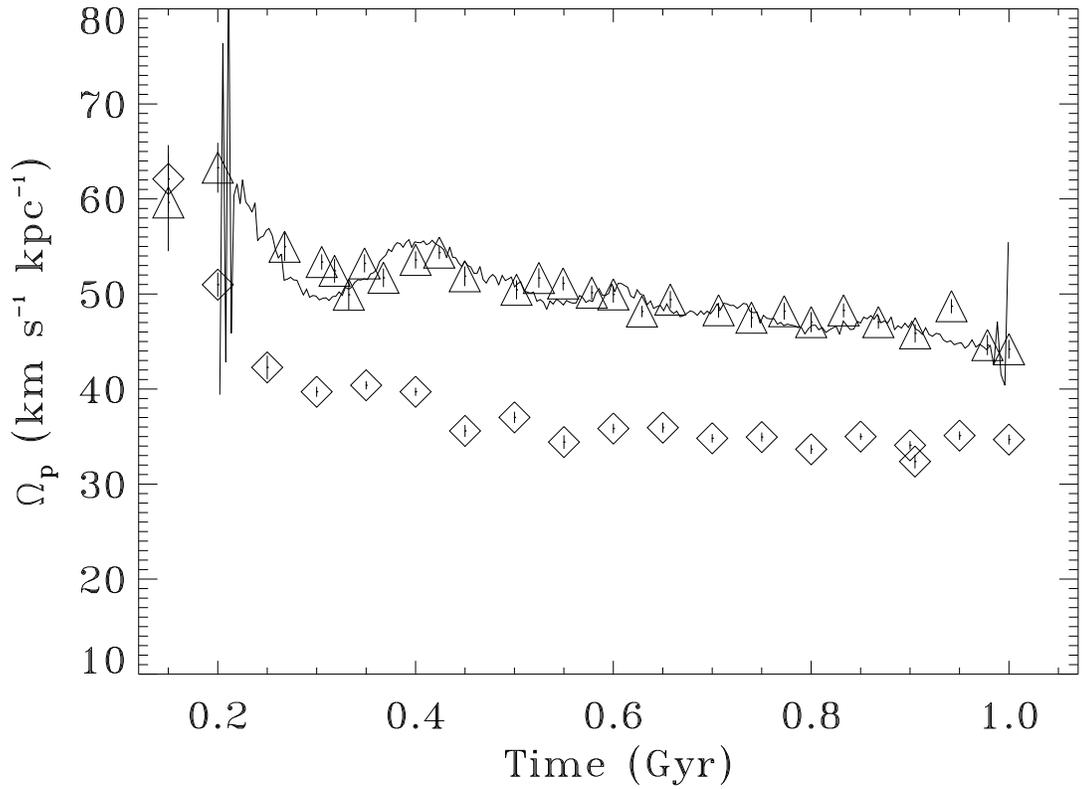}
\end{center}
\caption[Stellar pattern speeds $\Omega_p^s$]{\begin{small}Stellar pattern speeds $\Omega_p^s$ measured with the
automated TW method using the stellar velocity field of \FBDA\
(open triangles) and \FBDB\ (open diamonds). 1$-\sigma$ errors are overplotted as
vertical lines but are often smaller than the symbols. The pattern
speed computed during the simulation is plotted as a continuous
line.\end{small}} \label{fig:A}
\end{figure}

First, we have tested our method on the stellar velocity field of
\FBDA\ (cf. Fig.~\ref{fig:A}). The values of $\Omega_p^s$ are very
similar to those directly computed with the bar position-angles
determined every Myr during the computation of the simulations. Thus,
we confirm that the TW method is a very efficient tool to determine the bar
pattern speed in numerical simulations.
%
%

In the case of \FBDB, a pure N-body simulation without gas and star
formation but with the same total \emph{initial} mass than \FBDA, the
bar rotates slower than for \FBDA. This effect is obviously due to a
completely different evolution of mass transfers. Indeed, for \FBDA,
the gas inflow towards the center is quick enough to sustain the bar
pattern speed at a high value when the bar appears ($\sim$~0.2~Gyr for
\FBDA). The new stellar population created during the gas inflow
remains inside the corotation, increasing the stellar mass.
%
%
\begin{figure}
\begin{center}
\plotone{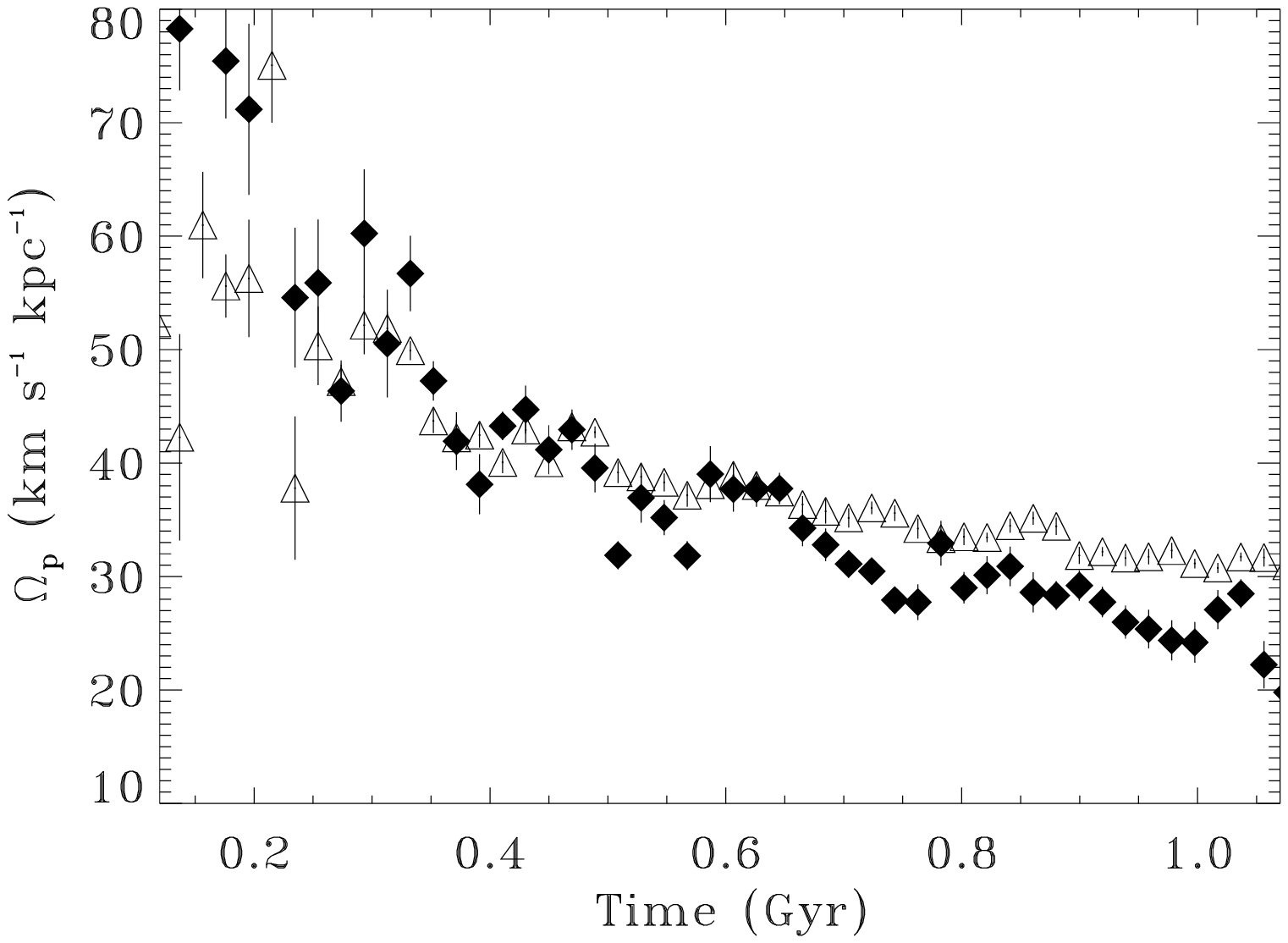}
\end{center}
 \caption[Gaz and stellar pattern speeds]{\begin{small}
Pattern speeds determined using the stellar velocity field
($\Omega_p^s$ open triangles) and the gas one ($\Omega_p^g$ full
diamonds) for \FBDG. 1$-\sigma$ errors are overplotted as vertical
lines.\end{small}} 
\label{fig:B}
\end{figure}
As for the observations (cf. Sect.~\ref{sec:newobs}), 
the velocity field of the gaseous component was also used to compute
$<V_{LOS,Y}>$. Fig.~\ref{fig:B} shows the results obtained with
\FBDG\ . For this run, star formation was switched off.  Once the bar
had settled in ($t \ga 0.25$~Gyr), and strong shocks had faded
away, $\Omega_p^s$ and $\Omega_p^g$ give similar estimations of
the pattern speed. Indeed, in absence of strong shocks, the gas
velocity field is close to the stellar one, at least in the
central region of the bar. Thus, on real data, it is mandatory to
avoid to include any region of strong shocks in the computation of
$<V_{LOS,Y}>$.

For $t \ga 0.6$~Gyr, $\Omega_p^g$ gives lower values than
$\Omega_p^s$. This bias can be removed by a careful selection of
the regions where the LOS velocities and mass densities are
measured. Indeed, a careful inspection of the projected mass
density and velocity fields shows that the gaseous bar could be
sometimes significantly shorter than the stellar one. Since the
automated TW method inconveniently selects some regions outside
the gaseous bar, this leads to an underestimation of the bar
pattern speed. Thus, the application of the TW method to
observational gas velocity fields needs a careful selection of “slit”, parallel to the major axis in the $<V_{LOS,Y}>$-$<X_Y>$ fit. The idea of the method is not to mask parts of the galaxy but to remove points from the $<V_{LOS,Y}>$-$<X_Y>$ fit using the curve  $<Y_X>$-$<X_Y>$ to break down to degeneracy of the $<V_{LOS,Y}>$-$<X_Y>$ fit. In that way, parts of the galaxy are NEVER masked and we NEVER violate the basic assumption of the method. In order to clearly select the regions of the galaxy used to apply the TW to determine one pattern speed, one must select the correct slits along which the integral from minus infinity to infinity (or at least  from one border of the field to the other) will be done to respect the basic assumption of the TW method. 

\begin{figure}
\begin{center}
\plotone{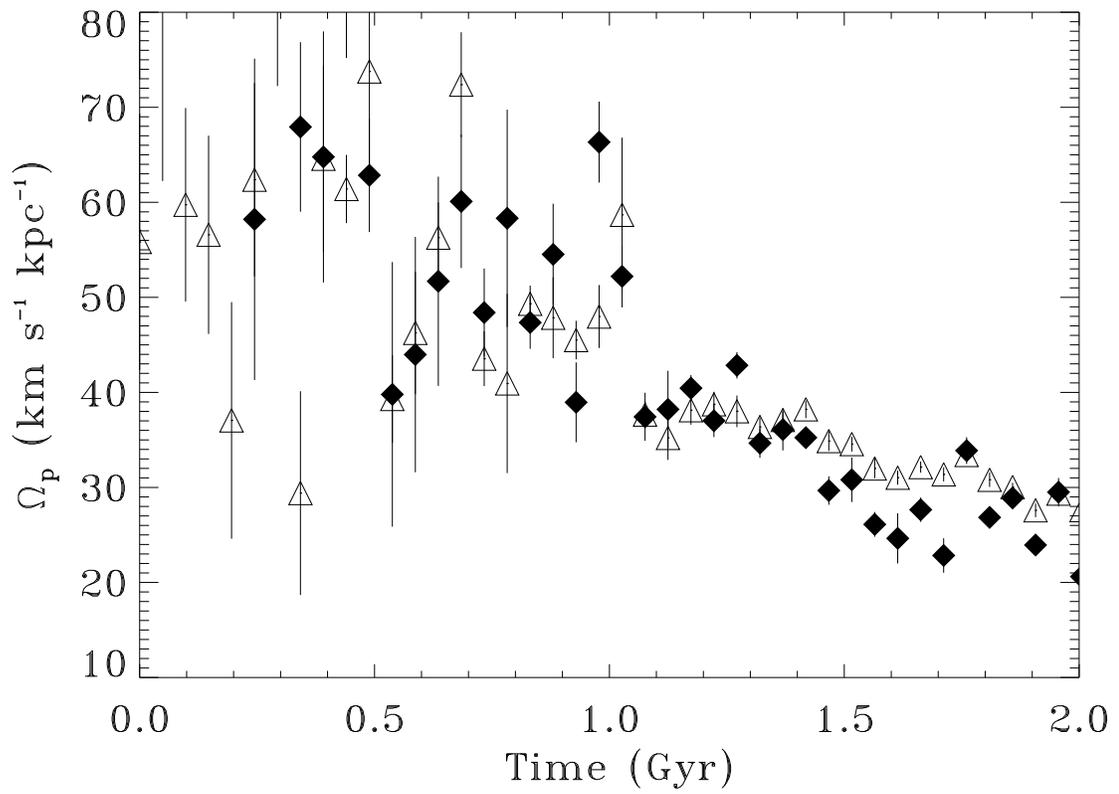}
\end{center}
\caption{As
Fig.~\ref{fig:B} for \EACB. Star formation is inhibited.}
\label{fig:D}
\end{figure}

\begin{figure}
\begin{center}
\plotone{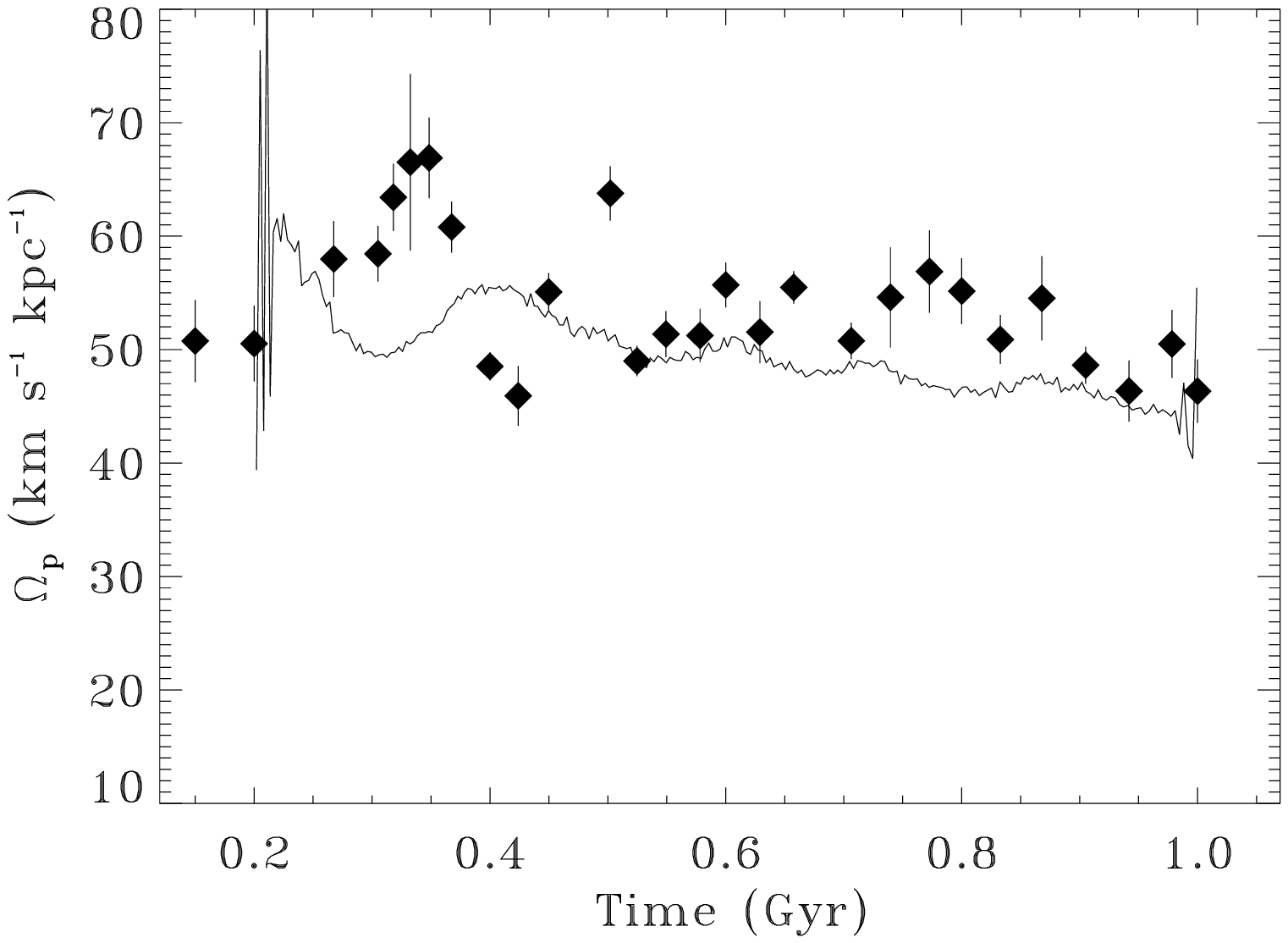}
\end{center}
\caption[TW using the velocity field and mass density of the
gaseous component]{\begin{small}As
Fig.~\ref{fig:B} using the velocity field and mass density of the
gaseous component for \FBDA. $\Omega_p^s$ is not plotted again as it closely follows the pattern speed measured during the simulation.\end{small}} \label{fig:C}
\end{figure}

A living dark halo does not change significantly the results. As
expected, the gravitational interaction made by a low mass dark halo
(\EAAJ) does not change much the bar pattern speed. In the case of
\EACB\ (Fig.~\ref{fig:D}), once the bar is formed, $\Omega_p^g$ can be
a good estimator of $\Omega_p$ on condition that the gaseous inner
ring is excluded from the measurements. Otherwise, $\Omega_p^g$ is
biased towards lower values.
%
%
When star formation is switched on, the TW method applied to the
gaseous component is less accurate (Fig. \ref{fig:C}). 
Indeed, $\Omega_p^g$
slightly overestimates the real bar pattern speed, especially when the
SFR is high ($ \ga 5$~\sfrunit). The reason for this behaviour is
indirectly linked to the star formation process. Indeed, the main
source of errors is the presence of strong and persistent gradient in
the gaseous velocity fields. These regions of high density are
obviously the privileged sites of stellar formation. Since the mass
distribution is completely different from \FBDG\ because of the new
stellar population, shocks regularly occur, especially in the inner
region of the bar where the gas inflow is stopped. Finally, when most
of the gas has been consumed by star formation (i.e. $t \ga 0.9$~Gyr), its
velocity field tends to be like the stellar one, leading to better
estimates of $\Omega$.

However, the discrepancies between $\Omega_p^s$ and $\Omega_p^g$
remain within $\pm 10$~\kms~kpc$^{-1}$. Thus the TW method could be
applied to the gaseous velocity fields to get a rough
estimated of the bar pattern speed, under the condition that regions
of shocks are avoided (slits passing through those regions are not taken into account in the final $<V_{LOS,Y}>$-$<X_Y>$ fit) and measurements are confined to regions where the gaseous bar is well formed.
%
%
All our results have been checked with \FBDT, the same simulation than
\FBDA\ but with 5 times more particles. The SFR history is different
(cf. Fig.~\ref{fig:A}) from \FBDA\ but we reach the same conclusions.
We have also applied the same method as the one used with real data
(cf. Sect.~\ref{sec:newobs}). Such method uses an adaptive spatial
smoothing algorithm based on the Voronoi tesselation to produce maps
of mass and velocity fields. We do not show the results here since
they are very similar to the automated TW method.

\section{The Case of M100 (NGC 4321)}

\subsection{General description of M100}

\begin{figure}
\plotone{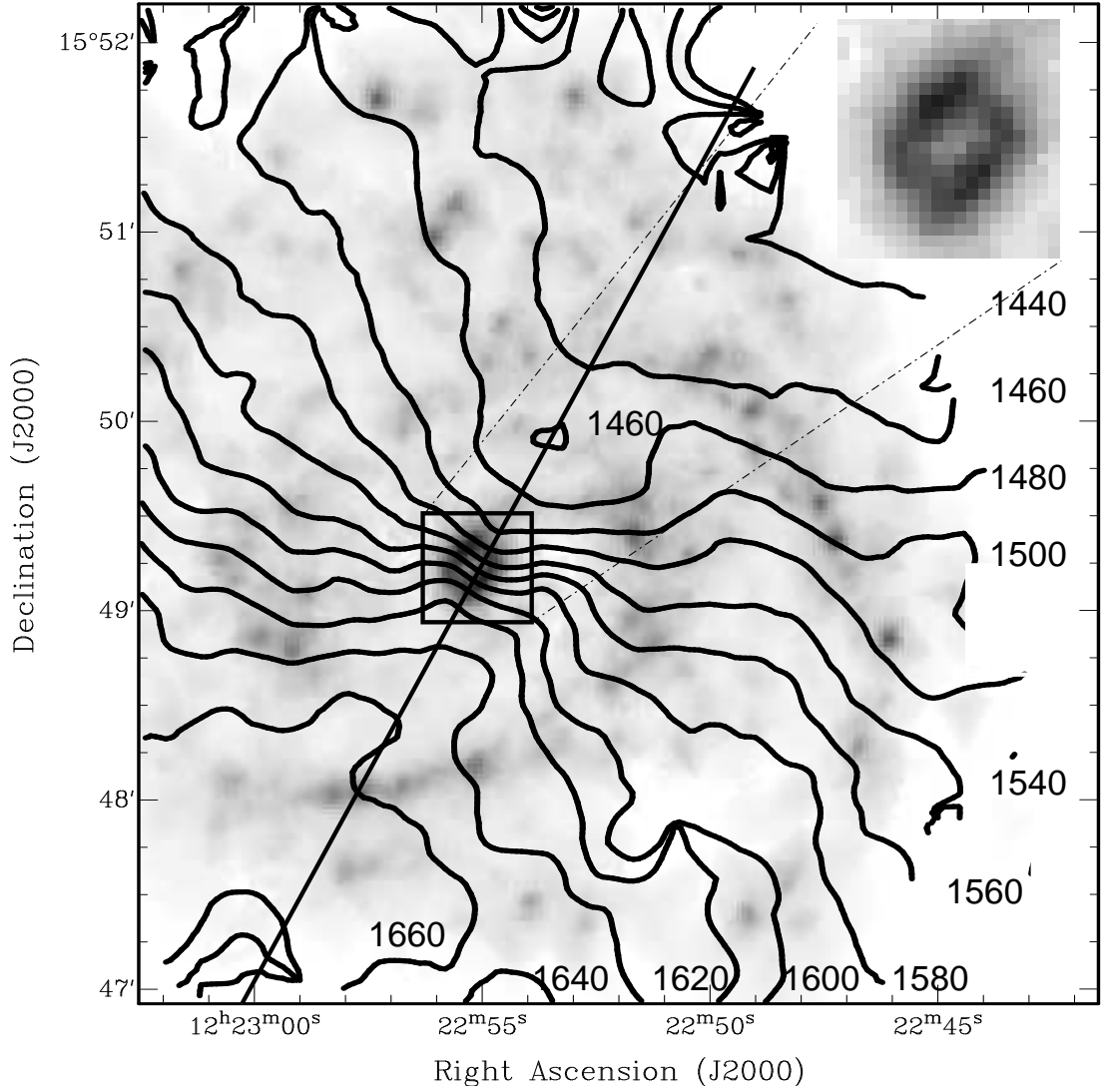}
\caption[M100]{M100: Isovelocity contours superimposed
on the \ha monochromatic image of the galaxy. The observations were obtained on the 1.6m of the Observatoire du mont M\'egantic with \FM. The label of the
isovelocities are given in \kms.  The major axis of the galaxy is
indicated by a thick line. The top-right zoom is an enlargement
of the central region of the galaxy.}
\label{iso} 
\end{figure}

M100 (NGC 4321, VCC 596) is one of the most studied object among
the nearby  barred spirals. The main parameters of M100 can be
found in Table \ref{tabM100paras} and its \ha velocity field is shown in Figure \ref{iso},
superimposed on the \ha monochromatic image. This grand-design
SAB(s)bc galaxy lies in the Virgo galaxy cluster, projected at
$\sim 1.1$ Mpc from M87 in the cluster core and has two apparent
dwarf companions, VCC 608 and VCC 634, which are only at 24 and 28
kpc respectively (in projection) from its nucleus. M100 shows a
small bulge, two prominent symmetric spiral arms lying within a
more complex spiral structure and two diffuse stellar extensions
to the North and South. The northern extension appears to end
close to VCC 608. Classical dust lanes are observed throughout the
whole disk along the spiral arms and the leading sides of the weak
bar.

%
%
\begin{table*}
\caption[Main parameters of M100]{Main Parameters of M100 (NGC4321).}
\begin{center}
\begin{tabular}[t]{lll}
\hline\hline \noalign{\medskip}
$\alpha$ (J2000)        &  12$^{h}$ 22$^{m}$ 54.9$^{s}$  & \\
$\delta$ (J2000)      & +15\Deg ~49\Min ~21\Sec & \\
Morphological type & SAB(s)bc & (1)\\
Heliocentric radial velocity        & 1586$\pm3$ \kms & (1)\\
Adopted Distance        & 16.1 Mpc (78 pc/arcsec) & (2)\\
Isophotal major diameter, $D_{25}$\Min  & 7.4\Min$\pm$0.2\Min (34.6$\pm$0.9 kpc) & (1)\\
Exponential disk scale length (K band), $\alpha^{-1}$ & 59.7\Sec$\pm$2.2\Sec (4.6$\pm$0.2 kpc) & (3)\\
Mean axis ratio  & 0.85$\pm$0.04 & (1)\\
Inclination, i     & 31.7\Deg$\pm$0.7\Deg & (4)\\
Position angle, PA & 27.0\Deg$\pm$1.0\Deg & (4)\\
Total apparent magnitude, B$_T$(0) & 9.98 & (1)\\
Absolute magnitude, M$_B$  & -21.05 & (2)\\
\noalign{\medskip} \hline
\end{tabular}
\end{center}
$^{(1)}$ RC3 data (de Vaucouleurs et al. 1991); $^{(2)}$
Cepheid-based distance from Ferrarese et al.(1996); $^{(3)}$
2MASS K-band ellipse fitting; $^{(4)}$ Based on the velocity field of Fig 6, for more 
details see the \BB sample kinematics form Hernandez et al. 2005.
\label{tabM100paras}
\end{table*}
%
%

The galaxy has been frequently mapped in the \ha emission-line using
high-resolution Fabry-Perot interferometry
(Arsenault, Roy \& Boulesteix 1990; Cepa \& Beckman 1990;
Knapen et al. 1995; Canzian \& Allen 1997; Knapen et al. 2000),
in the molecular CO emission-line (Canzian 1990,
Sakamoto et al. 1995,  Rand 1995,
Garcia-Burillo et al. 1998, Helfer et al. 2003)
and in the 21-cm HI emission-line (Cayatte et al. 1990, Knapen et al. 1993).
The HI disk is almost totally confined within the optical one but with
a slight lopsideness towards the SW (Knapen et al. 1993).
This asymmetry  could be either due to ram pressure stripping by the
Virgo intra-cluster medium  or to a tidal perturbation from a companion.
The \hI, CO and \ha velocity fields show kinematic disturbances such
as streaming motions along the spiral arms and  a
central S-shape distortion of the iso-velocity contours along the bar axis.

The circumnuclear region (CNR) of M100 has particularly received a
great deal of attention in multi-wavelength observations
because of the presence of an enhanced star formation region as a
four-armed \ha ring-like  structure and a CO spiral-like structure
(e.g. Sakamoto et al. 1995; Knapen et al. 2000). Both   the
ionized and molecular gas components  extend up to $R \sim
20$\Sec\ and show a central peak of emission centered on the
galactic nucleus. However, the location of gas intensity maxima in
the \ha ring and  in the CO nuclear spiral arms do not coincide
with the \ha peaks located more inward than those of the CO
emission. A lower limit to the CNR total molecular mass is $\sim
2\times 10^9$ \msol, as inferred from the CO integrated intensity
(Sakamoto et al. 1995). The kinematics of the \ha nuclear ring
shows non-circular motions and the \ha RC steeply rises in the
innermost 2\Sec\ (160 pc), reaching a  velocity
 of 150 \kms\ (Knapen et al. 2000).

The origin of the CNR SF-region is supposed to be due to gas
accumulation in the vicinity of the inner Lindblad resonances of
the disk  with density waves induced by the bar potential (Knapen
et al. 1995; Sakamoto et al. 1995; Sempere et al. 1995). Hence,
the determination of the  pattern speed $\Omega_p$ of the
large-scale bar of M100 (or reciprocally of the corotation radius)
has become a crucial imperative in order to understand the role of
resonances in the evolution of the galaxy. A value of $\Omega_p$
within the range $\sim$ 20-40 \kms kpc$^{-1}$ has thus emerged
from several observational methods and theoretical models (see
Table \ref{table4}).

The presence or absence of a secondary bar in the nuclear region
remains an open question. On one hand, some numerical simulations
need two nested bars rotating at significantly different
pattern-speeds to explain the CNR gaseous morphology
(Garcia-Burillo et al. 1998). In this framework, the secondary
nuclear bar rotates about seven times faster than the large-scale
primary bar, having its corotation region near the inner Lindblad
resonances of the primary bar. However, other simulations require
a nuclear bar that would rotate only three times faster than the
large scale one (Wada, Sakamoto \& Minezaki 1998). On the other
hand, Knapen et al. (2000) claimed that only a single $R \sim
60$\Sec\ large-scale bar is observed in M100 by means of numerical
models and isophotal analysis of NIR images which show an almost
perfect alignment of the isophotes in the innermost and external
parts of the bar. Their simulations predict that the nuclear
structure, which was formed from a single-barred potential,
corotates with the bar. Hence, the use of the TW method should in
principle help to resolve this question.

\subsection{New Fabry-Perot Observations.}

%
%
\begin{table*}
\caption[Journal of Fabry Perot observations]{Journal of Fabry Perot observations.} 
\begin{flushleft}
\begin{tabular}[t]{lll}
\hline\hline \noalign{\medskip}

Telescope        &  Observatoire du mont M\'egantic & 1.6 m\\
Equipment        & \FM\,@ Cassegrain &\\
Calibration      & Neon Comparison light & $\lambda$ 6598.95 \AA \\
Interference filter   & Central wavelength& $\lambda$ 6605 \AA \\
                      & FWHM              & 15 \AA \\
                      & Transmission at maximum & 0.75 \\
              & Temperature during the observations & $-$25$^o$C \\
Date             & & 2003, February,  25 \\
Exposure time    & Total & 260 minutes \\
                 & Elementary & 15 secondes\\
                 & Per channel & 5 minutes\\
Detector         & IPCS & GaAs tube  \\
Perot--Fabry & Interference Order & 899 @ 6562.78 \AA  \\
         & Free Spectral Range at \ha & 333.36~km~s$^{-1}$ \\
         & \textit{Finesse}$^{(1)}$ at \ha & 23\\
         & Spectral resolution at \ha & 20~677$^{(2)}$ \\
Sampling & Number of Scanning Steps & 52\\
         & Sampling Step & 0.14~\AA\ (16~km~s$^{-1}$)\\
         & Total Field & 824"$\times$824" \\
 & & (512$\times$512 px$^2$)$^{(3)}$ \\
         & Pixel Size & 1.61" (0.126 kpc) \\
         & Seeing     & $\sim$1.42" \\
\noalign{\medskip} \hline
\end{tabular}
\end{flushleft}
$^{(1)}$ Mean \textit{Finesse} through the field of view\\
$^{(2)}$ For a signal to noise ratio of 5 at the sample step\\
$^{(3)}$ After binning 2*2, the original GaAs system
providing 1024$\times $1024 px$^2$ 
\label{tabletwoh3}
\end{table*}

New observations of M100 were obtained in February 2003 with the
Fabry Perot instrument \FM\footnote{\FM\,(for \textbf{Fa}bry-Perot de \textbf{N}ouvelle
\textbf{T}echnologie pour l'\textbf{O}bservatoire du \textbf{m}ont
\textbf{M}\'egantic) was developed by the Laboratoire
d'Astrophysique Exp\'erimentale (LAE, Montr\'eal), \texttt{http://www.astro.umontreal.ca/fantomm}} (Hernandez et al. 2003)
on the mont M\'egantic 
Observatory (OMM) 1.6m telescope in the frame of a new
large observational program of barred galaxies (\BB sample: Hernandez,
et al. 2005).
\FM\, is composed of a focal reducer (bringing the original f/8
focal ratio of the Cassegrain focus to f/2), a scanning
Fabry-Perot and an Image Photon Counting System (IPCS)
based on a new technology GaAs amplifier tube which has a
high quantum efficiency (Gach et al. 2002).
The journal of the observations and the
observational set up is given in Table \ref{tabletwoh3}.

The reduction of the data cubes was performed using the package
ADHOCw (Boulesteix, 2004) rewritten with large improvements under
the IDL package. The
signal measured along the scanning sequence was separated into two
parts: (1) an almost constant level produced by the continuum
light in a narrow passband around \ha (image not shown) and (2) a
varying part produced by the \ha line (referred hereafter as the
monochromatic map).

In order to increase the signal-to-noise ratio, an adaptive spatial
smoothing based on the Voronoi tesselations method (Daigle et al. 2005b, Cappellari \& Copin 2002) and applied to the 3D data cubes (Daigle et al. 2005b) was used to produce the monochromatic images and the velocity fields. The strong OH night sky lines passing through the filter were reconstructed into a cube and subtracted from the galaxy's spectrum. The wavelength calibration was obtained by scanning the narrow Ne 6599 \AA\ line under the same conditions as the observations. The velocities measured relative to the systemic velocity are very accurate, with an error of a fraction of a channel width (${\rm <~3 \,~km~s^{-1}}$) over the whole field. Details on the observations and on the data reduction will be given in forthcoming papers (Hernandez et al. 2005; Daigle et al. 2005b).

The position angle (PA), inclination, systemic velocity and rotation center
(X$_{cen}$, Y$_{cen}$) have
been calculated using \textsc{ROTCUR} provided by the \textsc{Gipsy}
package and \textsc{Karma} to construct the Position-Velocity plot
(Hernandez et al. 2005). The fit was performed using a robust $\chi^2$
model and the central regions of the velocity field were 
masked to avoid contamination due to the bar. Debattista
(2003) has shown that the TW measurements of bar pattern speeds are
sensitive to errors in the PA of the disc. 2D
velocity fields of extended galaxies allow an accurate determination
of the PA, reducing the errors when using the TW method.

\subsection{Results.}
\label{sec:newobs}
%
%
\begin{figure*}
\begin{center}
\plottwo{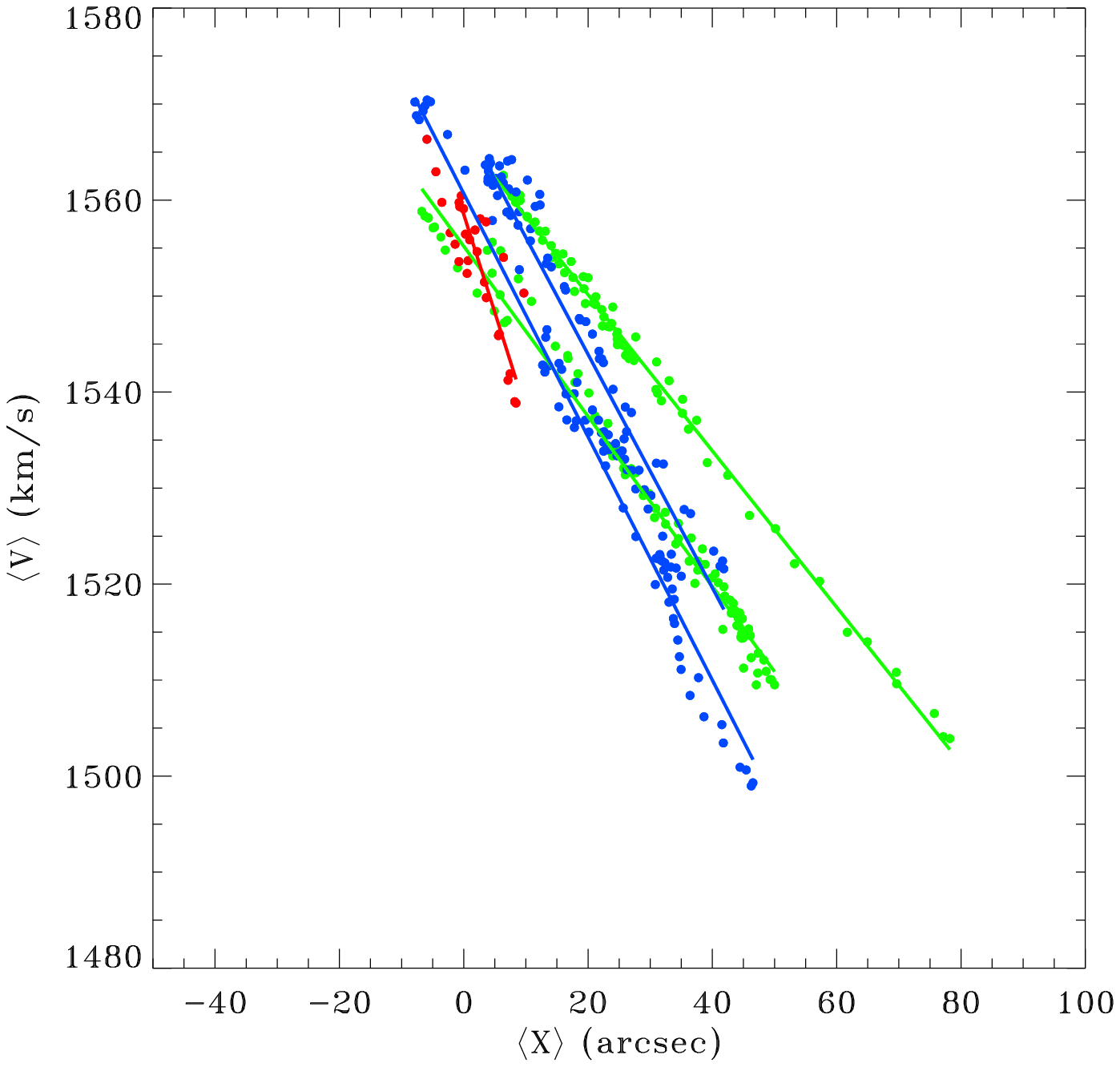}{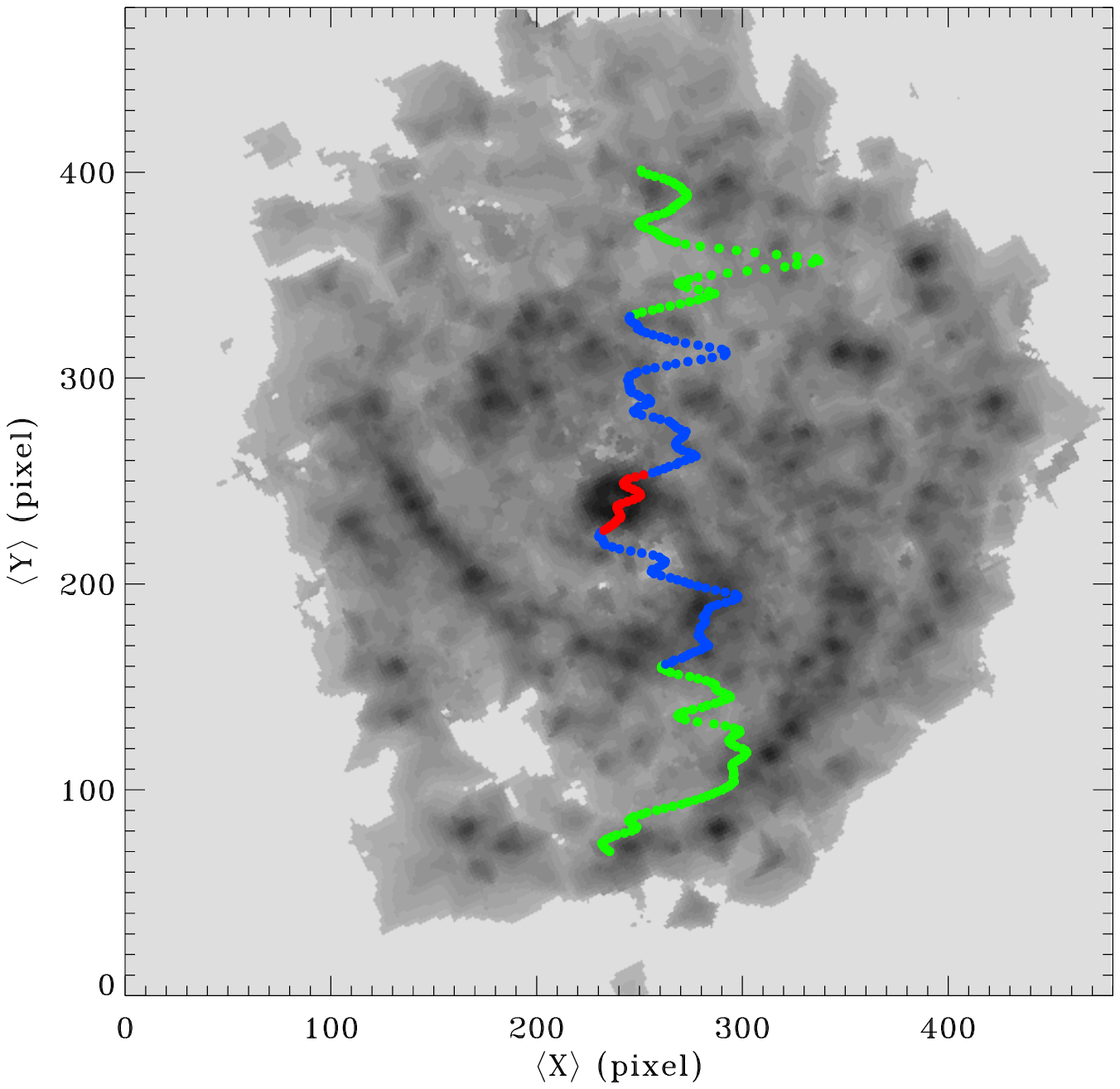}
\end{center}
\caption[TW method applied to M100]{(left) Mean line-of-sight
velocity versus mean position for M100. Both quantities are averaged along
the same strip parallel to the major axis. The degeneracy
introduced during the integration along the X-axis may be followed
with the colors using the righthandside plot. Strait lines
represent the linear fits. (right) Mean Y-position versus
mean X-position. Both quantities are averaged along the same strip
parallel to the major axis. The red are for the nuclear
structure, blue for the bar region and the green for
the disk spiral.}
\label{hatwm100} 
\end{figure*}
In Figure \ref{hatwm100} (left), the intensity-weighted 
line-of-sight velocity $<V_{LOS,Y}>$
is plotted versus the intensity-weighted position coordinate,
$<X_Y>$. The \ha velocity field and the \ha monochromatic image have been used to measure $<V_{LOS,Y}>$ and $<X_Y>$. The degeneracy introduced during the calculation of $<X_Y>$
(integration along the X-axis) may be followed on
Figure \ref{hatwm100} (right) which gives the position of $<X_Y>$. For instance,
the red dots in the center of the galaxy correspond to $<Y_X>$
located near the nuclear structure. The blue ones stresses the
bar region, and the green ones the disk structure.

Corsini, Debattista \& Aguerri (2003) and Rand \& Wallin (2004) show that the TW method can be used to determine multiple $\Omega_p$ . If more than one pattern is present for a given aperture (as in e.g., the bar-within-bar models of Friedli \& Martinet 1993), $\Omega_p$ will have contributions from the various patterns.  Both $<X_Y>$,$<Y_X>$ and $<V_{LOS,Y}>$-$<X_Y>$ plots are used to select the appropriate “slits” with regard to the galaxy region to clearly identify the pattern of the selected wave. A robust fit is then done on each series of points to calculate the pattern speed of each waves.  In the case of M100, Figure 7 (left) shows the distribution of points when the TW is used over the galaxy with slits parallel to the major axis and from +/- infinity (i.e. in the limits of the field of view) while Figure 7 (right) shows the $<X_Y>$,$<Y_X>$ plot superimposed with the \ha image to help us disentangle clearly the different pattern speeds. Using the assumptions of section 3.2 to select the correct slit to calculate the appropriate pattern speed, a robust $\chi^2$ linear fit is then performed on the three series of points (blue, green, red).

$\Omega_p$ of the bar is found to be 30.3$\pm$1.8 \kms kpc$^{-1}$.
The distance has been chosen such that we can compare with
$\Omega_p$ from other sources. The error calculation takes into
account the error on the inclination, on the mean line of sight
velocity, on the mean position along the line of nodes and on the
PA. The errors are computed from the tilted ring models. Table \ref{table4}
presents the results of the TW method applied to M100. All the pattern speeds are
scaled to our inclination of 31.7\Deg\ and distance of 16.1 Mpc.

\section{Discussion.}

\subsection{Multiple Pattern Speeds.}

The pattern speed of the spiral arms may differ from that of the
bar (e.g. Sellwood \& Sparke 1988; Sellwood 1993; Rautiainen \&
Salo 1999). Different scenarii may explain the connection between
bar and spiral arms: \emph{(a)} corotating bar and spiral arms;
\emph{(b)} independent bar and spiral arms possessing different
pattern speeds connected by non-linear mode coupling.

\emph{(a)} In some cases, the bar and the spiral structure are
clearly corotating: \emph{(i)} the spiral arms start from the ends
of the bar (e.g. NGC 1365); \emph{(ii)} the fraction of grand
design spirals is higher in early type barred galaxies as compared
to non-barred ones (this may be the case also for late type barred
galaxies, e.g. NGC 157); \emph{(iii)} the size of the two armed
spiral in galaxies correlates with the size of the bar (e.g.
Elmegreen \& Elmegreen 1989, 1995); \emph{(iv)} the response to an
analytic bar potential is a bar driven spiral (e.g. Sanders \&
Huntley 1976); \emph{(v)} the outer rings usually correspond to
the outer Lindblad resonance in barred galaxies (e.g. Buta 1995).
Considering this scenario \emph{(a)}, it is hard to explain
\emph{(i)} multi-armed barred galaxies (e.g. ESO 566-24);
\emph{(ii)} spiral arms which do not start from the ends of the
bar; \emph{(iii)} flocculent barred galaxies (Buta 1995);
\emph{(iv)} the absence of rings in many barred galaxies (Sellwood
\& Wilkinson 1993).  These observations may be explained if bars
and spiral arms are either independent features or non-linearly
coupled as it could be the case for instance in galaxies like NGC
1068, NGC 1398, NGC 1566, NGC 2273.

\emph{(b)} Bars and spiral arms may be independent features and
have different pattern speeds (e.g. simulations by Sellwood 1985;
Sellwood \& Sparke 1988). Bar and spiral have different pattern speeds
connected by a non-linear mode coupling (Tagger et al. 1987;
Masset \& Tagger 1997; Rautiainen \& Salo 1999). In this
scenario, the corotation of the bar and the inner Lindblad
resonance of the spiral overlap in radius, which results in a
 transfer of energy and angular momentum between the modes. Several
simultaneous spiral modes can also coexist in the disk, even
overlapping in radius. These galaxies have inner spiral
(corotating with the bar) and outer spiral with a separate and lower pattern speed (Rautiainen \& Salo 1999).  On the other
hand, mode coupling may be stronger when the halo contribution to the
rotation curve is large (e.g. Debattista \& Sellwood 1998;
Rautiainen \& Salo 1999). Gnedin, Goodman \& Frei (1995) have shown that a gravitational angular-momentum flux, or torque, can be measured directly from the mass distribution in spirals.  These authors have carried out this measurement for M100 and concluded that the spiral structure seen in M100 may not be typical of its past or its future. These torques depend not on the pattern speed or permanence of the arms but only on the non-axisymmetric mass distribution.

In addition to the main bar component, many nuclear bars have been
observed (e.g. Buta \& Crocker 1993; Wozniak et al. 1995; Friedli
et al. 1996). At least some of these small scale bars have higher
pattern speeds than the main bar (Friedli \& Martinet 1993) and
inner bar formed before the main bar (Rautiainen \& Salo 1999).
In the case of M100, the nuclear bar has a fast pattern
($\Omega_p$ =160~\kms~$kpc^{-1}$; Garcia-Burillo et al. 1998) and is decoupled from the slow
pattern of the outer bar+spiral ($\Omega_p$=23~\kms~$kpc^{-1}$;
Garcia-Burillo et al., 1998). Solutions based on a single pattern
hypothesis for the whole disk cannot fit the observed molecular
gas response and fail to account for the relation between other
stellar and gaseous tracers (Garcia-Burillo et al. 1998).

From deep surface photometry in the K band obtained for 54 normal
spiral galaxies, Grosb{\o}l, Patsis \& Pompei (2004) found in several cases
that bars are significantly offseted compared to the starting points
of the main spiral pattern. This indicates that bar and spiral
have different pattern speeds.

\subsection{M100: Comparison with other studies.}

The number of barred galaxies that have been observed to date
using the TW method, mainly SB0 galaxies, is too small to
ascertain unequivocally whether centrally concentrated dark matter
haloes are truly absent in barred galaxies.

\begin{itemize}
\item In the stellar dynamical theory, the spiral arm amplitudes oscillate because of 
differential crowding near and between wave-orbit resonances. Three cycles of such 
oscillations have been found in B and I-band by Elmegreen \& Elmegreen (1989).  
Using $R_{25}=D_{25}/2=$3\Min.42 (de Vaucouleur et al. 1976), these authors supposed 
that the inner gap located at $0.35\times R_{25}$ is the "$+4:1$" resonance. Thus, 
power law extrapolation of the RC of M100 (extracted from Rubin et al. 1980, 
where $V(R)=r^{\alpha}$ with $\alpha=0.1$ for Rubin et al. instead of $\alpha=0.35$, 
for this present study) locates the corotation at $0.6\times R_{25}$, the OLR 
at $1.1\times R_{25}$ and the ILR at $0.13\times R_{25}$.  With a corotation 
at $0.6\times R_{25}$, the pattern speed is 21\kms~kpc$^{-1}$ (scaled to the 
same distance of 16.1 Mpc). 

\item The corotation resonance has been found within the range 101-128 arcsec (see Table \ref{tabreso}) from the \ha kinematics of the gas (Canzian et al. 1993).

\item Garcia-Burillo et al. (1994) discussed two different methods to measure the pattern speeds. Firstly, using CO observations, they were seeking for the detection of the change of sign of the radial streaming motions, as predicted by the theory, when going beyond the corotation. They found no change of sign in the radial streaming motions. Therefore, according to the observed kinematics, they inferred that the whole inner spiral structure is located inside corotation. Secondly, they made numerical simulations of the molecular cloud hydrodynamics and compared the gas response with the spiral structure seen in the optical and CO observations. Their best fit solution lead to $\Omega_p =$ 25\kms~kpc$^{-1}$ (scaled to our distance and inclination), implying that corotation lies at a radius approximately equal to 110 arcsec. This value, based on a global fit of the spiral using numerical simulations, gives a much more trusty determination for the corotation but is in clear contradiction with the observational determination quoted above that places corotation at the outer disk.

\item  Two different methods to derive the pattern speed have also been used by Sempere 
et al. (1995). The first method, based on the change of sign of the radial streaming motions 
beyond the corotation (Canzian 1993) lead to $\Omega_p =$ 25\kms~kpc$^{-1}$, that locates 
corotation in the middle of the disc (8-11 kpc i.e. 82\Arc-113\Arc).  The second method, 
involving hydrodynamic numerical simulations of the molecular cloud in a potential derived 
from an R-band image of the galaxy lead to $\Omega_p$ $\simeq$ 25\kms~kpc$^{-1}$. 
This validates the picture where the stellar bar ends within the corotation and the outer 
spiral lies outside the corotation. 

\item Rand (1995) estimates a value lower than 37\kms~kpc$^{-1}$ by identifying the CR with the location where no tangential streaming is observed when the CO arm crosses the major axis.

\item Wada et al. (1998) compared CO observations by Sakamoto et al. (1995) with a two-dimensional hydrodynamical and analytical bar model. Their best model agrees well with Knapen et al. (2000) about the double ILR, and with M100 having a single stellar bar with a pattern speed of 69\kms~kpc$^{-1}$ (in excellent agreement with 70\kms~kpc$^{-1}$ found in Knapen et al.).

\item Garc�-Burillo et al. (1998) claimed that two bars rotate at different angular speeds. 
The present study indicates clearly that the inner structure has a different pattern speed 
from the bar but it can not determinate precisely the pattern speed of the secondary bar.

\item Knapen et al. (2000) studied the circumnuclear starburst region of M100 and concluded 
that both morphology and kinematics require the presence of a double inner Lindblad resonance 
in order to explain the observed twisting of the near-infrared isophotes and the gas velocity 
field. The results of Knapen et al. (2000) are different from those 
of Garc�-Burillo et al.(1998).

\item Using this empirical relationship and deprojected bar, Sheth et al. (2002) measured a bar pattern speed of 35\kms~kpc$^{-1}$ on the CO rotation curve of M100 by Das et al. (2001).

\item Rand \& Wallin (2004) applied the TW method of pattern speed determination to CO emission (Sempere \& Garcia-Burillo 1997).  They assumed this galaxy is molecule-dominated and found that the method is insensitive to the bar pattern speed because the bar is nearly parallel to the major axis. They found a spiral pattern speed of 28$\pm5$ \kms~kpc$^{-1}$. Nevertheless, these authors found that the spiral pattern speed found agrees with previous estimates of the bar pattern speed, suggesting that these two structures are parts of a single pattern.

\item Corsini et al (2004) measured the bar pattern speed using the TW method.  
They compared the value with recent high-resolution N-body simulations of bars in 
cosmologically-motivated dark matter halos (Valenzuela \& Klypin 2003), and concluded that the bars are not located inside centrally concentrated halos and that N-body models 
produce slower bars than observed.  We found the corotation of M100 at the radius 
$R_{CR}$ (r$=$94\Arc) and the outer Lindblad resonance $OLR$ (r$=$145\Arc).  
If we conjecture that the end of the bar and the OLR match, thus M100 is in the 
forbidden area of their plot.
\end{itemize} 

Table \ref{tabreso} presents the location of resonances for all the previous resonances of the bar. The comparison with the present study is very consistent, especially for the bar corotation radius ($CR$).

\begin{figure*}
\plotone{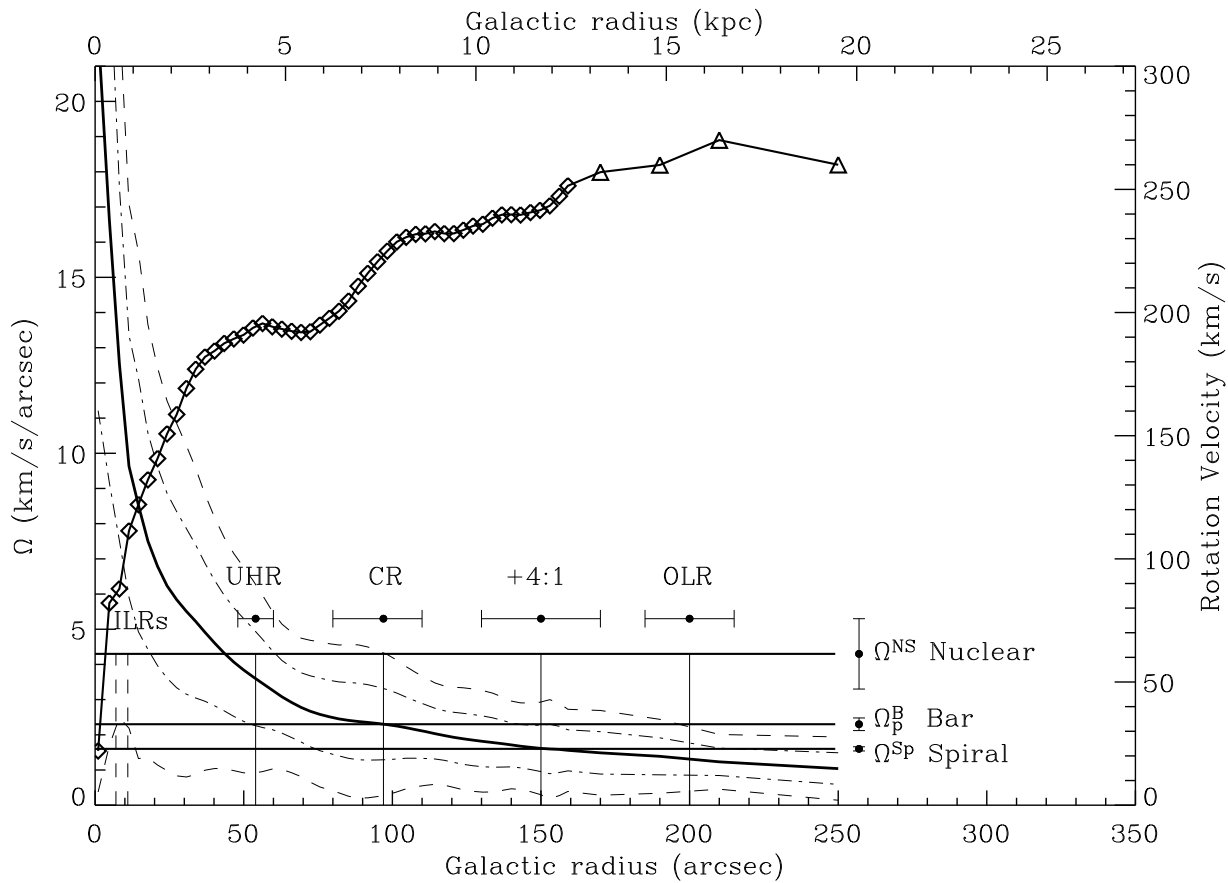}
\caption[RC of M100]{M100. The rotation curve of the galaxy is
drawn by the thin continuous line and the velocity scale is on the
right of the plot. \ha data are represented by losanges and are used for galactic 
radii $\leq 159\Arc$. HI data (Knapen et al. 1993) 
are represented by triangles and are only used for galactic radii $\geq 160\Arc$, due to 
their poor spatial resolution ($\simeq$45\Arc). The other curves are respectively, from the top
upper one to the bottom one, the $\Omega+\kappa/2$ (dash),
$\Omega+\kappa/4$ (dash-dot), $\Omega$ (thick continuous),
$\Omega-\kappa/4$ (dash-dot) and $\Omega-\kappa/2$ (dash) curves.
The expected positions of the bar resonances for $\Omega_P^B=$2.3\kms
arcsec$^{-1}$ are indicated by the vertical thin 
lines - surrounded by dark thick fill dots inclunding errors bars - respectively the two 
inner resonance $ILR_1$ (r$=$ 0.9 kpc),
$ILR_2$ (r$=$0.5 kpc); the UHR resonance (r$=$4.2 kpc); the corotation 
radius $R_{CR}$ (r$=$7.4 kpc), the '$+$4:1' resonance (r$=$11.3 kpc), and the 
OLR (r$=$15.22kpc). The values of $\Omega$s for the 3 patterns are indicated on 
the righthand side of the figure, with the appropriate vertical errors bars. See text 
for more details.}
\label{Fig4} 
\end{figure*}

Figure \ref{Fig4} presents the rotation curve of M100, the curve $\Omega(R)$,
where $R$ is the galactic radius, with the position of the resonances
using \ha\ data obtained with \FM\ and HI data extracted from Knapen
et al. (1993). The other curves are respectively, from the top to the
bottom, the $\Omega+\kappa/2$ (dash), $\Omega+\kappa/4$ (dash-dot),
$\Omega$ (thick continuous), $\Omega-\kappa/4$ (dash-dot) and
$\Omega-\kappa/2$ (dash) curves. The solid horizontal lines represent
respectively, from the top to the bottom, $\Omega^{NS}$ from the
nuclear structure, $\Omega_p^B$ from the bar and $\Omega^{Sp}$ from
the spiral pattern derived from the TW method using \has. $1\pm\sigma$
errors, on the three $\Omega$ are drawn in the lower righthandside
of the graph. For the bar, these errors are reported in terms of radii
to determine the range of resonance radii.

Fig.~\ref{Fig4} gives a clear evidence that the '$+$4:1' resonance of the bar,
located in the middle of the disc, is very close to the corotation of
the spiral. Beyond this radius ($\approx 11-12$ kpc) the spiral arms
vanish. This is a rather unexpected result since, in the case of a
single pattern speed, it has been shown that \textbf{1)} a massive
self-gravitating spiral lies between the ILR and the UHR and \textbf{2)} in
presence of a bar which fixes the pattern speed, a perturbative
(i.e. non self-gravitating) spiral structure lies between the
corotation and the OLR. However, all these results have been obtained
with the linear theory of density waves.

Since the original simulations of Sellwood (1985) and the theoretical
explanations of Tagger et al. (1987) and Sygnet et al. (1988), the
non-linear coupling of density waves has been recognized as an
efficient coupling mechanism between large scale morphological
features such as bars and spiral arms. In most cases of non-linear
coupling reported so far, waves are coupled thanks to the coincidence
of the bar corotation and the spiral ILR. The coupling of two $m=2$
modes generates two beat waves of modes $m=0$ and $m=4$. It has been
showed that such coincidence of resonances is the most efficient
configuration for energy and momentum transfers because the beat waves
also have a Lindblad resonance at the same radius.  The coupling
between a '$+$4:1' resonance and a corotation is not forbidden even if it
seems to be a less favourable configuration.  The theory of non-linear
coupling allows however the existence of others kinds of coupling:
Rautinainen \& Salo (1999) reported a case of coupling between the bar
corotation and the spiral ultra harmonic resonance (UHR or '$-$4:1').
%
%

If we disregard the error on $\Omega^{IS}$ of the structure inside
the circum-nuclear ring, we could be tempted to see another resonances
overlap between the OLR of the nuclear structure and the bar
corotation. This location is indeed associated with an abrupt change
of the pitch angle of the spiral arms. However, the inaccurate
location of the resonances due to the error on $\Omega^{IS}$
prevents us to draw any definite conclusion on a possible coupling.

Another noteworthy property must be emphasized: the three different
pattern speeds could be related by $ \Omega^{Sp} + \Omega_p^{B}
\approx \Omega_p^{IS} $. The non-linear coupling of two density waves
predict such a relationship between the two initial waves and the beat
waves. But, in our case, the nuclear structure cannot be considered as
a beat wave resulting from the interaction of the bar and the spiral
structure since it has neither the right location nor the right
azimuthal wave number $m$. Thus, a complete understanding of M100
certainly needs the development of a model based of the non-linear
coupling of three density waves, which is outside the scope of this
paper.
%
%
\begin{table*}
\caption[Location of the resonances of M100]{Location of the resonance radii in \textit{arcsec} for the bar. This study vs litterature scaled to our inclination and distance.\\}
\begin{tabular}{l|llllll}
\hline \hline 
ref. & OLR & $+4:1$ & \textbf{CR} & UHR & ILR1 & ILR2 \\
\hline
This study & 200 & 150 & 97 & 54 & 11 & 7 \\
~~~~~~[range in \Arc] & \small[185:205] & \small[130:167] & \small[80:110] & \small[48:60] & n/a & n/a \\
\small Elmegreen et al. (1992) & & & 118 & & & \\
\small Elmegreen et al. (1989) & 225 & n/a & 123 & 71 & n/a & n/a \\ 
\small Canzian (1993) & & & \small 114-[101:128] & & \\
\small Canzian et al. (1997) & & & \small 98-[88:108]& & \\
\small Sempere et al. (1995) & & & \small 97-[82:113] & & \\
\small Garcia-Burillo et al. (1994) & & & 110 & & \\
\hline
\end{tabular}
\label{tabreso}
\end{table*}
%
%
\begin{table*}
\caption[M100: Comparison of pattern speeds]{M100: Comparison of pattern speeds}
\begin{tabular}[t]{lllll}
\hline \hline\noalign{\medskip}
Method                 &Spectral    &\multicolumn{3}{c}{Pattern Speed}\\
                       &Range       &$\Omega_p^{(1)}$                 &$\Omega_p^{(2)}$ &Ref.\\
                       &            &(\emph{\kms} $arcsec^{-1}$)      &(\emph{\kms} $kpc^{-1}$)&\\
\hline
TW$^{(a)}$ Nuclear Struct.&\ha       &4.3$\pm$1.0   &   55$\pm$5       &$^{(b)}$\\
TW Bar                 &\ha         &2.30$\pm$0.18 &   30.3$\pm$1.9   &$^{(b)}$\\
TW Spiral pattern      &\ha         &1.6$\pm$0.06  &   20.4$\pm$0.8   &$^{(b)}$\\
\hline
TW                     &CO          &2.53          &   32.4          & $^{(c)}$\\
Canzian, 1993          &\ha$^{(d)}$ &1.92          &   19.8          &$^{(d)}$\\
Resonance 4:1          &B\&I-band   &1.73          &   17.9          &$^{(e)}$\\
\hline
\multicolumn{2}{l}{N-body$^{(f)}$ }                &1.92          &   19.8 &$^{(d)}$\\
\multicolumn{2}{l}{SPH-Hydrodynamical}             &1.89          &   22.8 &$^{(g)}$\\
\multicolumn{2}{l}{N-body simulation}              &3.36          &   40.5 &$^{(h)}$\\
\hline
\end{tabular}
$~~~~~~~~~~~~~~~~~~~~~~~~~~~~~~~~~~~~~~~~~~~~~~~~~~~~~$ \\
$^{(1)}$ Scaled to our inclination of 31.7\Deg; $^{(2)}$ Scaled to
our inclination of 31.7\Deg and distance of 16.1 Mpc.\\
$^{(a)}$ Tremaine-Weinberg 1984; $^{(b)}$ Present work; $^{(c)}$
Rand and Wallin 2004; $^{(d)}$ Sempere et al. 1995; $^{(e)}$
Elmegreen \& Elmegreen 1989; $^{(f)}$ Combes \& G\'erin 1985; $^{(g)}$
Wada et al. 1998; $^{(h)}$ Rand 1995. 
\label{table4}
\end{table*}

\label{sec:rc}

\section{Conclusions.}

High spectral and spatial resolution \ha monochromatic image and \ha velocity field have been 
presented in order to study the multiple pattern speeds of M100 using the Tremaine-Weinberg 
method. At the same time, the TW method has been tested on various numerical 
simulations to test its relevance. The main conclusions are the following: 

\begin{itemize}
\item The TW method can be applied to the gaseous velocity fields to get the bar pattern speed, under the condition that regions
of shocks are avoided and measurements are confined to regions where the gaseous bar is well formed.	
\item The application of the TW method to gas velocity fields needs a careful selection of
the region included in the $<V_{LOS,Y}>$-$<X_Y>$ fit.
\item A living dark halo does not change significantly the results. 
\item When star formation is switched on, the TW method applied to the
gaseous component is less accurate. The main source of errors is the presence of strong and persistent gradient in
the gaseous velocity fields. 
\item The TW method provides clear information about multiple pattern speeds.
\item M100 has been analyzed using the TW method. Three pattern speeds are clearly seen and measured with a good precision compared to other studies. Errors on the PA and the inclination have been minimized due to the nature of the two dimensional velocity maps, improving the relevance of the results obtained for the three pattern speeds.
\item A coupling between resonance the '$+$4:1' of the bar and the corotation of the spiral is found and are in agreement with the theory of non linear coupling of spiral modes.  The coupling is not forbidden even if it seems to be a less favourable configuration. The theory of non-linear coupling allows however the existence of others kinds of coupling as reported by Rautinainen \& Salo (1999).
\item A restriction for a more accurate determination of the different radii and resonnaces comes from the intrinsic nature of M100. The central bar distorts the spiral arm structure and the corotation is not a narrow, but a relatively extended region.

\end{itemize}

\clearpage

\end{document}